\documentstyle[epsf,epsfig,preprint,aps,amsmath,amssymb]{revtex}
\draft \preprint{SNUTP 03-010}
\begin{document}
\title{\Large\bf Three family 
$Z_3$ orbifold trinification, MSSM and doublet-triplet 
splitting problem}
\author{Kang-Sin Choi\footnote{ugha@phya.snu.ac.kr} and Jihn E.
Kim\footnote{jekim@phyp.snu.ac.kr}}
\address{
School of Physics and Center for Theoretical Physics,\\ 
Seoul National University, Seoul 151-747,
Korea}\maketitle

\begin{abstract}
A $Z_3$ orbifold compactification of $E_8\times E_8^\prime$ heterotic 
string is considered toward a trinification $SU(3)^3$ with three
light families. The GUT scale VEV's of the $SU(2)_W\times 
U(1)_Y\times SU(3)_c$ singlet chiral fields in two
sets of the trinification spectrum allow
an acceptable symmetry breaking pattern toward MSSM. We show that
a doublet-triplet splitting is related to the absence of a
$\Delta B$ nonzero operator. 
\vskip 0.5cm\noindent [Key words: $Z_3$ orbifold, superstring, 
trinification, three families]
\end{abstract}


\newcommand{\bea}{\begin{eqnarray}}
\newcommand{\eea}{\end{eqnarray}}
\def\beq{\begin{equation}}
\def\eeq{\end{equation}}

\def\one{\bf 1}
\def\two{\bf 2}
\def\five{\bf 5}
\def\ten{\bf 10}
\def\tenb{\overline{\bf 10}}
\def\fiveb{\overline{\bf 5}}
\def\threeb{{\bf\overline{3}}}
\def\three{{\bf 3}}
\def\fb{{\overline{F}\,}}
\def\hb{{\overline{h}}}
\def\Hb{{\overline{H}\,}}

\def\sw{$\sin^2\theta_W$}
\def\MG{$M_{GUT}$}

\def\slash#1{#1\!\!\!\!\!\!/}
\def\hf{\frac12}

\def\A{{\cal A}}
\def\Q{{\cal Q}}

\def\b3{\bar 3}

\newcommand{\debug}{\emph{!!! CHECK !!!}}

\newcommand{\dd}{\mathrm{d}\,}
\newcommand{\Tr}{\mathrm{Tr}}
\newcommand{\drep}[2]{(\mathbf{#1},\mathbf{#2})}

\newpage

\section{Introduction}

It seems that the family structure of the standard model(SM)
is completed with three light ones. This observation
stems from the recent experiments toward understanding
neutrino oscillation, Big Bang nucleosynthesis, and 
experiments saturating the unitarity triangle. For a long
time, the question, $\lq\lq$Why are there three
light families?", has been the heart of the family problem.
In 4 dimensional(4D) field theories, the
grand unification idea with
a big gauge group was suggested toward this family structure, 
which is called the
{\it grand unification of families(GUF)}\cite{georgi}.
For the GUF idea to work from a bottom-up approach,
the three different gauge coupling constants observed 
at the electroweak scale should meet at a 
grand unification(GUT) scale \MG. With the three light families and 
one Higgs doublet scalar fields, they do not meet. 
But one can make them meet by introducing a number
of particles beyond the three family structure of the SM. 
One interesting possibility is the particle spectrum of  
the minimal supersymmetric SM(MSSM)\cite{gaugeunif}.

With the advent of superstring models, the GUF idea seems to
be automatically implemented. In particular, the 
10 dimensional(10D) heterotic string models need big gauge groups,
$E_8\times E_8^\prime$ or $SO(32)$\cite{ghmr}. 
Among these, the $E_8\times
E_8^\prime$ has attracted a particular attention.
However, the big gauge group
is given in 10D, and one has to hide six internal spaces
to contact with our 4D world. This process of hiding six 
internal spaces is known as $\lq\lq$compactification", 
accompanies the breaking of the big 10D gauge group, and also 
generates multi-families in 4D\cite{candelas,dixon}. 
The most serious objective
in this compactification has been to obtain the MSSM in 4D. 
For an $N=1$ supersymmetry, the internal space with an
$SU(3)$ holonomy has been suggested first~\cite{candelas}.
But a more interesting and easily soluble case is the
orbifold compactification~\cite{dixon}. In particular,
the $Z_3$ orbifold models with two Wilson lines attracted
a great deal of attention because of the multiplicity 3
in the spectrum~\cite{iknq}. Along this line, 
the {\it standard-like models}, which allow three families
and $SU(3)_c\times SU(2)\times U(1)^n$ groups, have been extensively 
studied~\cite{iknq,imnq}.

The standard-like models, however, suffered from the following
two problems:\\
\indent ($i$) the $\sin^2\theta_W$ problem, and\\
\indent ($ii$) the problem of too many Higgs doublets.

\noindent
With the MSSM spectrum, it is necessary to assume that the
unification value of \sw\ is $\frac38$ to reconcile with
the low energy data on $\alpha_{QCD}, \alpha_{em}$ and
\sw. 
The $\sin^2\theta_W$ problem ($i$) is that it is generally difficult
to obtain $\frac38$ for the unification value of \sw.
The problem of too many Higgs doublets is that
the standard-like models have many pairs of Higgs doublets 
while the MSSM needs just one pair. To solve the above 
problems, recently it was 
suggested to unify the standard model in a semi-simple gauge 
group at the compactification scale so that the electroweak 
hypercharge is not leaked to $U(1)^n$ factors\cite{kim03}.
In \cite{kim03}, the motivation has been to 
embed the electroweak hypercharge in semi-simple groups 
with no need for the adjoint representation(HESSNA).
In the HESSNA, the QCD gauge group must be already factored out
so that an adjoint representation is not needed.
The simplest HESSNA is the $SU(3)^3$ gauge group with the
so-called {\it trinification}\cite{glashow} spectrum for one family,
\begin{equation}\label{trinification}
{\bf (\bar 3,3,1)+(1,\bar 3,3)+(3,1,\bar 3)}.
\end{equation}

This leads us to search for simple
$SU(3)^3$ models for HESSNA. In this paper, we present a
$Z_3$ orbifold model which leads to a model close
to the MSSM below a GUT scale.
We also show a correlation between the doublet-triplet splitting
and the $\Delta B$ nonzero operator $u^cd^cd^{\prime c}$.

\section{A $Z_3$ orbifold model from $E_8\times E_8^\prime$}

The heterotic string theory has $N=4$ supersymmetry 
from the 4D viewpoint. To obtain chiral fermions in 4D, 
we have to reduce $N=4$ supersymmetry down to $N=1$. 
The $Z_3$ orbifold reduces $N=4$ down to $N=1$ when we compactify
the six internal spaces\cite{dixon}. The six internal
spaces are split into a direct product of three two-dimensional
tori($y_1-y_2;y_3-y_4;y_5-y_6$). A $Z_3$ orbifolding 
of two dimensional torus gives three
fixed points; thus three $Z_3$ orbifolded tori have 27
fixed points. The 27 fixed points are not distinguishable
unless one introduces Wilson lines. The shift vector $V$
and the six Wilsone lines $a_i(i=1,\cdots,6)$
are embedded in the gauge group $E_8\times E_8^\prime$. 
The $a_1$ is transformed to $a_2$ by a $Z_3$ transformation, and 
we consider only three independent Wilson lines:
$a_1=a_2,a_3=a_4,a_5=a_6$\cite{inq}.

The model we study here is\footnote{A precurser of the present
model with $V$ and $a_1$ was already given
before~\cite{kim03,kschk}.}
\begin{eqnarray}
&V = (0~ 0~ 0~ 0~ 0~ \frac13~ \frac13~ \frac23) (0~ 0~ 0~ 0~ 
0~ \frac13~ \frac13~ \frac23)\nonumber\\
&a_1 =  (\frac13~ \frac13~ \frac13~ 0~ 0~ \frac13~ \frac13~ 0) 
(\frac13~ \frac13~ 0~ 0~ 0~ 0~ \frac13~ \frac13)\label{model}\\
&a_3 =  (0~ 0~ 0~ 0~ 0~ 0~ 0~ 0~) (0~ 0~ \frac13~ \frac13~ 
\frac23~ 0~ 0~ 0)\nonumber
\end{eqnarray}
with $a_5=(0~\cdots)(0~\cdots)$. Eq. (\ref{model}) is allowed
in superstring orbifolds. For the conditions to be
satisfied, see Ref.~\cite{inq}.
The unbroken gauge group is $[SU(3)^3\times U(1)^2]\ \times\ 
[SU(3)^2\times U(1)^4]^\prime$.
Here, however, 
we assume that six $U(1)$'s are broken by VEV's of $SU(3)^5$
singlet fields at the string scale. Below the string scale, the 
effective gauge group is $SU(3)^3\times[SU(3)^2]^\prime$, and
hence the invariance under the nonabelian gauge group is our
main concern in this paper. In HESSNA, one does not have to know
the extra $U(1)$ quantum numbers to pinpoint the electroweak
hypercharge.

Thus in the observable sector, this compactification leads at 
low energy to an $N=1$ effective field theory $SU(3)^3$ 
with three copies of trinification spectrum (\ref{trinification}).
The massless chiral fields are presented in
Table I with the well-known method~\cite{inq,iknq,kim03}.  
Because there are nine twisted sectors, the
multiplicity in one twisted sector is 3. Because of $Z_3$, the chiral
fields of the untwisted sector also have the multiplicity 3.
These are the bases for three chiral families.
Note that the fields in the nine twisted sectors
of Table I form vectorlike
representations which can be removed at a GUT scale. Therefore,
we will be interested in the 3 copies of the trinification
spectrum appearing in the untwisted sector.

In many aspects for low energy physics, it is similar to an $E_6$ model 
with three families of {\bf 27}. In the present model, however, the
electroweak gauge group and $SU(3)_c$ are already split and we do not
need an adjoint representation for the symmetry breaking\cite{kim03}.
 
When one blows up the fixed points and obtain a smooth
Calabi-Yau manifold with an $SU(3)$ holonomy,
one $SU(3)$ factor from the orbifold is identified with the
$SU(3)$ holonomy and is removed from the low energy gauge 
group~\cite{dixon}. We can identify one of $SU(3)$'s 
in the hidden sector for this purpose if we wish. 

\section{The minimal supersymmetric standard model}

To obtain the low energy effective theory MSSM, we must break the
$SU(3)^3$ gauge symmetry down to the MSSM group 
$SU(2)_W\times U(1)_Y\times SU(3)_c$ at a GUT scale \MG.
Let us represent the trinification fields
of (\ref{trinification}) as
\begin{eqnarray}
{\bf (\bar 3,3,1)}\longrightarrow 
  \Psi_{[l=(\bar M,I,0)]}&=& \Psi_{(\bar 1,i,0)}(H_d)_{-\frac12}+ 
  \Psi_{(\bar 2,i,0)}(H_u)_{+\frac12}
  + \Psi_{(\bar 3,i,0)}(l)_{-\frac12}\nonumber\\
  &+&\Psi_{(\bar 1,3,0)}(N_5)_0+ 
  \Psi_{(\bar 2,3,0)}(e^+)_{+1}+ \Psi_{(\bar 3,3,0)}(N_{10})_0
\label{lepton}
\\
{\bf (1,\bar 3,3)}\longrightarrow  
  \Psi_{[q=(0,\bar I,\alpha)]}\ &=& \Psi_{(0, \bar i,\alpha)}
  (q)_{+\frac16}+ \Psi_{(0,\bar 3,\alpha)}(D)_{-\frac13}
\label{quark}
\\
{\bf (3,1,\bar 3)}\longrightarrow  
  \Psi_{[a=(M,0,\bar\alpha)]}&=& \Psi_{(1,0,\bar\alpha)}
  (d^c)_{\frac13}+ \Psi_{(2,0,\bar\alpha)}(u^c)_{-\frac23}
  + \Psi_{(3,0,\bar\alpha)}(\overline{D})_{+\frac13}.
\label{antiquark}
\end{eqnarray}
where $M,I,\alpha$ are the $SU(3)_1, SU(3)_2\equiv SU(3)_W,$ 
and $SU(3)_3\equiv SU(3)_c$ indices. Under the SM gauge group,
$I=i=\{1,2\}$ and $\alpha=\{red,green,blue\}$ represent $SU(2)_W$ and 
$SU(3)_c$ indices, and we appropriately represented the
well-known notations for the SM fields in the 
parenthesis. The $U(1)_Y$ charges are shown
with subscripts. Let us call the three representations given 
in (\ref{lepton}), (\ref{quark}) and (\ref{antiquark}), as three
different {\it humors} and name them as {\it lepton--, quark--,}
and {\it antiquark--humors} because leptons, doublet quarks, and 
$u^c,d^c$ quarks appear there. In (\ref{lepton}) there are two fields 
which are neutral under the SM gauge group: $N_5$ and $N_{10}$.
Therefore,
GUT scale vacuum expectation values of these fields break
down the $SU(3)^3$ gauge group down to the SM gauge group,
\begin{eqnarray}
SU(3)^3&\xrightarrow[\langle N_{10}\rangle]& \ SU(2)_1\times
SU(2)_W\times U(1)_a\times SU(3)_c\nonumber\\
&\xrightarrow[\langle N_{5}^\prime\rangle]& 
\ SU(2)_W\times U(1)_Y\times SU(3)_c\label{step}
\end{eqnarray}

The symmetry breaking is achieved by giving VEV's to the
scalar partners of the three family trinification fields.
In the first step of symmetry breaking (\ref{step}), 
9 Goldstone bosons are absorbed through the Higgs mechanism 
to the gauge bosons. These are contained in
$H_d,H_u,l,N_5,e^+$, and $N_{10}$. In the second step of
(\ref{step}), 3 further Goldstone bosons are absorbed to
gauge bosons through the Higgs mechanism. The resulting gauge
group is the SM gauge group and must be anomaly free. 
The study of this symmetry breaking pattern is not trivial
and one must consider two steps of (\ref{step}) together.
With only one $(\bar 3,3,1)$ representation, 
we cannot break $SU(3)_1\times SU(3)_2$ down to $SU(2)_W
\times U(1)_Y$. We need at least two $(\bar 3,3,1)$ representations
which are supposed to be scalar partners of two out of three 
copies of (\ref{lepton}).
After the Higgs mechanism, the remaining SM fields are linear combinations
of the fields arising in (\ref{lepton}). Then we can redefine the
fields so that SM fields are renamed. The remaining fields from
two sets of (\ref{lepton}) must include two sets of \{$l_{-\frac12}, 
e^+$\}. If $H$ fields are removed, they must be vectorlike representations.
Otherwise, there appear anomalies. 
Note that Eqs. (\ref{quark}) and (\ref{antiquark}) lead to three quark
families, and hence the anomaly free condition dictates to have three
lepton families. Thus, after the Higgs mechanism there appear two
sets of \{$l_{-\frac12},e^+$\} from two sets of
$(\bar 3,3,1)$ for the spontaneous symmetry breaking. 
These $l_{-\frac12}$'s are the renamed
fields from the linear combinations of the original fields 
$H_d^{(1)},H_d^{(2)},l^{(1)}_{-\frac12}$, and $l^{(2)}_{-\frac12}$.

To discuss the light spectrum more concisely, let us utilize
the $N=1$ supersymmetry explicitly. Possible cubic
terms among the untwisted sector fields are~\cite{fiqs},
\begin{equation}\label{Yukawa}
{-\cal L}_Y=\frac{1}{3!} f_{abc}\Psi^a \Psi^b \Psi^c
\end{equation} 
where $a,b,c$ are the family indices. Note that we consider
only the $SU(3)^3$ symmetry. [At the fundamental level,
$f_{abc}$ are the coupling constant times ratios of
singlet VEV's to the string scale.] Note
that $f_{abc}$ is {\it completely symmetric}.
To distinguish the third family from the first two families 
participating in the GUT symmetry breaking, 
we postulate that $f_{ab3}=0$ if $a$ or
$b$ is in \{1,2\}. Therefore, let us study the GUT symmetry
breaking sector with $a,b,c\subset \{1,2\}$ first.
Assigning VEV's as $
\langle \Psi^{(1)}_{(\bar 3,3,0)}\rangle=\tilde V_1,\  
\langle \Psi^{(2)}_{(\bar 1,3,0)}\rangle=\tilde V_2,
$
we note that one $H_u$ and one combination $H_d^\prime$ (composed
of $H_d$'s and $l_{-\frac12}$'s) form Dirac particles at a GUT
scale. Therefore, out of 18 chiral fields we can figure out
ten fields first: four from massive $H_u$ and $H_d^\prime$ 
and six from two sets of $\{l_{-\frac12},e^+\}$. Thus, we 
can identify 12 Goldstone bosons among the remaining 8 complex(or
16 real) scalar fields. After the Higgs
mechanism(removing 12 real fields), the remaining
fields are two complex fields: $N_5$ and $N_{10}$. If we consider 
$SU(3)^3$ singlets $S$'s with GUT scale VEV's, these singlet 
neutrinos can obtain large masses. In this case,
we obtain only two sets of $\{l_{-\frac12},e^+\}$
from two sets of the trinification spectrum.
The third set of the trinification spectrum contains
one pair of $H_u$ and $H_d$ which is the needed
light Higgs doublet pair in the MSSM.

Out of the three sets of the trinification spectrum (\ref{trinification}),
thus we obtain three fermion families, and their superpartners. 
For the number of Higgs doublets, see below.

\section{Doublet-triplet splitting}

For the MSSM, we need a pair of Higgs doublets. But if the coupling
(\ref{Yukawa}) is completely general, we cannot achieve this objective
since $H_u$ and $H_d$ in the third family, not participating in the
GUT group breaking, will be heavy. We need a fine-tuning to
keep them light. But this fine-tuning is correlated with a 
$\Delta B\ne 0$ operator.

Before showing the doublet-triplet splitting
explicitly, we point out that the resolution of this 
doublet-triplet splitting problem in the flipped $SU(5)$
model~\cite{su51} heavily assumes the absence of $H_dH_u$ coupling.
It is the familiar $\mu$ problem, and can be solved by 
introducing a Peccei-Quinn symmetry\cite{kn}.
But in string theory, we can see that the $H_dH_u$ 
term cannot arise at the tree level. Since both $H_d$
and $H_u$ belong to (\ref{lepton}) in our compactification, 
a guessed term for $H_dH_u$, i.e. the term
among the light fields $(\bar 3,3,1)\cdot(\bar 3,3,1)$  
is forbidden from the gauge symmetry. 
In addition, however,
the coupling $(\bar 3,3,1)\cdot(\bar 3,3,1)\cdot(\bar 3,3,1)$
among the light fields, must be forbidden to remove
the $H_dH_u$ coupling at a GUT scale because $H_dH_u$ can arise
after giving a VEV to $N_5$ or $N_{10}$. Below we show that
this can be realized by a fine-tuning but this fine-tuning must be 
dictated from a $\Delta B$ nonzero operator.

The VEV's of $N_5$ and $N_{10}$ allow the following two types of nonvanishing 
mass terms. The first possibility is coming from $SU(3)^3$ singlets
by taking three different {\it humors}, and the second possibility is 
coming from $SU(3)^3$ singlets by picking up the same {\it humor} from 
$\Psi^a,\Psi^b,$ and $\Psi^c.$ In general, these two possibilities
are present. In the discussion on the GUT symmetry breaking, we
allowed both of these couplings. Below, we mainly focus on the couplings
of the third family. 

The first possibility gives masses to $D$ and $\overline{D}$.
For example, for $\langle N_{10}({\rm 3rd\ family})
\rangle=\tilde V_1$,\footnote{Before, we assigned VEV's only to
the first two families. Since we have figured out the light spectrum before
with two sets of (\ref{trinification}), now we can also assign
a VEV to the third family member. The composition of the new light
fields will be more complicated, but the number of light degrees will
be intact.} 
we obtain 
$ D M_D \overline{D}$
where 
$$
M_D=\tilde V_1\left(
\begin{matrix}
f_{113}\ \ f_{123}\ \ 0\ \\
f_{213}\ \ f_{223}\ \ 0\ \\
\ 0\ \ \ \ \ 0\ \ \ f_{333}
\end{matrix}
\right).
$$
Note that Det$M_D$ is nonzero, and three pairs of
$D$ and $\overline{D}$ are removed at a GUT scale.
Let us focus on the $f_{333}$ coupling below.

The second possibility allows a $u^cd^cd^{\prime c}$ coupling,
considering the {\it antiquark humor}. It violates the $R$-parity,
and is dangerous for proton stability. Therefore,
we choose a fine-tuning such that the second possibility from 
$f_{333}$ is excluded.

Let us try to implement a permutation symmetry $S_3$ in the
$SU(3)^3$ model for a simpler discussion of the couplings. 
The three humor sets (\ref{lepton}), (\ref{quark}), and
(\ref{antiquark}), i.e. {\it lepton--, quark--}, and {\it
antiquark--humors},
$\Psi_l,\Psi_q,$ and $\Psi_a$ are represented as a
singlet and a doublet of the permutation of $\{l,q,a\}$~\cite{segre},
\begin{eqnarray}
\Psi_0 &=& \frac{1}{\sqrt{3}}(\Psi_l+\Psi_q+\Psi_a)\nonumber\\
\Psi_+ &=& \frac{1}{\sqrt{3}}(\Psi_l+\omega\Psi_q+
         \omega^2\Psi_a)\label{permrep}\nonumber\\
\Psi_- &=& \frac{1}{\sqrt{3}}(\Psi_l+\bar\omega\Psi_q+
         \bar\omega^2\Psi_a)\nonumber
\end{eqnarray}
where $\omega$ and $\bar\omega$ are the cube roots
of unity $\omega=e^{2\pi i/3},\bar\omega=e^{4\pi i/3}$.
Note that $\Psi_0$ is a singlet under the permutation of
$l,q,a$. On the other hand $\Psi_\pm$ goes into a
multiple of $\Psi_\mp$. Thus, $\Psi_+$ and $\Psi_-$ form
a doublet under permutation, which we can represent as
$\Psi_{doublet}\equiv( \Psi_+, \Psi_-)^T$.
The $S_3$ invariant cubic couplings are $\Psi_0^3$ and 
$\Psi_0\Psi^+\Psi^-$. In terms of {\it humors}, these are
\begin{eqnarray}
\Psi_0^3 =\mbox{$\frac{1}{3\sqrt{3}}$}(&\Psi_l^3&+\Psi_q^3
          +\Psi_a^3+3\Psi_l^2\Psi_q+3\Psi_l^2\Psi_a
          +3\Psi_q^2\Psi_l\nonumber\\
         &+& 3\Psi_q^2\Psi_a+3\Psi_a^2\Psi_l+3\Psi_a^2\Psi_q
          +6\Psi_l\Psi_q\Psi_a)\label{singlet}\nonumber\\
\Psi_0\Psi_+\Psi_- = &\frac{1}{3\sqrt{3}}&(\Psi_l^3
          +\Psi_q^3+\Psi_a^3-3\Psi_l\Psi_q\Psi_a)\nonumber
\end{eqnarray}
The above
couplings include the so-called $R$-parity violating couplings
of the MSSM. In particular, the $\Delta B\ne 0$ operator 
$u^cd^cd^{\prime c}$(the so-called $\lambda^{\prime\prime}$ 
coupling) is dangerous.
It is contained in $\Psi_a^3$. To remove this $\Delta B\ne 0$
coupling $\Psi_a^3$, we fine-tune the $\Psi_0^3$ and $\Psi_0\Psi_+
\Psi_-$ couplings such that they have the same
magnitude but the opposite signs. Then, 
the $S_3$ invariant coupling is
\begin{eqnarray}
&\mbox{$\frac{1}{\sqrt{3}}$}(\Psi_l^2\Psi_q+\Psi_l^2\Psi_a
          +\Psi_q^2\Psi_l
         + \Psi_q^2\Psi_a+\Psi_a^2\Psi_l+\Psi_a^2\Psi_q
          +3\Psi_l\Psi_q\Psi_a)\nonumber\\
& \longrightarrow 
          \sqrt{3}\Psi_l\Psi_q\Psi_a
\end{eqnarray}
where in the second line we excluded the terms not 
allowed by the gauge invariance.
Thus, the phenomenological requirement for proton
stability excludes the
$H_dH_u$ allowing term $\Psi_l^3$(the second possibility), 
and hence $H_d$ and $H_u$ are left as light particles.
Furthermore, the coupling allows the first 
possibility, i.e. the coupling chooses different humors in
the cubic terms, and hence removes the color triplets
$D$ and $\overline{D}$, realizing the doublet-triplet splitting.

If this argument is applied to the first two families, we
will end up with two pairs of Higgs doublets, one pair too much.
We must remove one more pair, but then we must allow a 
$\lambda^{\prime\prime}$ coupling. A sizable 
$\lambda^{\prime\prime}$ for the $t$ quark family is not forbidden
very strongly phenomenologically(For proton decay, a product
$\lambda^\prime\lambda^{\prime\prime}$ is constrained.). To
obtain a phenomenologically acceptable MSSM, we may require this kind
of fine-tuning, forbidding the same humor coupling, among
the two lighter families\footnote{The previous discussion on
the GUT symmetry breaking assumed the same humor coupling,
mainly to find out the heavy fields, i.e. non-Goldstone fields.}; 
but allow an O(1) same humor coupling for the $t$ family.

\section{conclusion}

In conclusion, we constructed a $Z_3$ orbifold trinification 
model with three light families, and showed 
that the symmetry breaking leads to a spectrum close
to the MSSM. The discussion on keeping
one pair of $H_u$ and $H_d$ light needed a fine-tuning in this
paper, but this fine tuning has been shown to be
correlated with the absence of $\Delta B$
nonzero operator $u^cd^cd^{\prime c}$. It will be very interesting
if this fine-tuning is naturally obtained.

\acknowledgments
We thank Kyuwan Hwang for a useful discussion. 
We also thank the Physikalisches Institut of the
Universit\"at Bonn for the hospitality extended to us
during our visit when this work was initiated. This work
is supported in part by the KOSEF Sundo Grant(2002),
the BK21 program of Ministry of Education, and Korea
Research Foundation Grant No. KRF-PBRG-2002-070-C00022.

\begin{table}
\caption{\it Every representation has multiplicity
3 because of $Z_3$ for the case of U and two Wilson lines for
the cases of nine T's. 
}
\begin{tabular}{c|c}
\hline
sector & fields \\
\hline
U  & $(\b3,3,1)(1,1)+ (3,1,\b3)(1,1)+(1,\b3,3)(1,1)$\\
   &   $+3(1,1,1)(1,3)$ \\
T0 ($V$) & nine singlets \\
T1 ($V+a_1$) & $(1,3,1)(1,1)+(3,1,1)(1,1)+(1,1,3)(1,1)$ \\
T2 ($V-a_1$) & $(1,\b3,1)(1,1)+(\b3,1,1)(1,1)+(1,1,\b3)(1,1)$ \\
T3 ($V+a_3$) & nine singlets \\
T4 ($V-a_3$) & $3(1,1,1)(1,\b3)$ \\
T5 ($V+a_1+a_3$) & $(1,1,3)(1,1)+(3,1,1)(1,1)+(1,3,1)(1,1)$ \\
T6 ($V+a_1-a_3$) & $(1,1,3)(1,1)+(3,1,1)(1,1)+(1,3,1)(1,1)$ \\
T7 ($V-a_1+a_3$) & $(1,1,\b3)(1,1)+(\b3,1,1)(1,1)+(1,\b3,1)(1,1)$ \\
T8 ($V-a_1-a_3$) & $(1,1,\b3)(1,1)+(\b3,1,1)(1,1)+(1,\b3,1)(1,1)$ \\
\hline
\end{tabular}
\end{table}

\end{document}